# Sequential vibrational resonance in multistable systems

J H Yang[1,2] and X B Liu[1,*]

[1] Institute of Vibration Engineering Research, Nanjing University of Aeronautics and Astronautics, Nanjing 210016, P. R. China

[2] School of Mechanical and Electrical Engineering, China University of Mining and Technology, Xuzhou 221116, P. R. China

[*] Corresponding author. E-mail: xbliu@nuaa.edu.cn; Tel.: +86-25-84892106;

## Abstract

The phenomenon of sequential vibrational resonance existed in a multistable system that is excited by both high- and low-frequency signals is reported. By the method of direct separation of motions, the theoretical investigation on vibrational resonance is conducted in both cases of underdamped and overdamped, and the analytical predictions are in good agreement with the numerical simulations. In view of the theoretical results, the zero-order Bessel function related to the high-frequency signal is included in the renormalized resonant frequency, and which leads to the appearance of the sequential vibrational resonance. In the case of underdamped system with small damping coefficient, the resonance occurs in a series of discrete parameter regions. The sequential vibrational resonance is different from the traditional multiple vibrational resonance, because its appearance is much more regular. The results in this work may be helpful in the field of signal processing electronics, especially for dealing with the very-low-frequency signal.

# 1. Introduction

In the vibrational mechanics scheme, a fascinating phenomenon named vibrational resonance (VR) has attracted more and more attentions since it was first found and reported by Landa and McClintock [1]. VR indicates that the weak low-frequency signal can be amplified by adjusting another excitation-including high-frequency signal. By reason of the significant applications of biharmonic signals in a variety of fields such as brain dynamics [2, 3], laser physics [4], acoustics [5], *etc.*, so far, via numerical, analytical and experimental methods, VR has been widely investigated by many researchers in the domains of bistable systems [6-11], optical systems [12, 13], coupled anharmonic oscillators [14, 15], noisy structure [16], neuron systems [17-19], complex networks [20] and quintic oscillators [21, 22]. Very recently, VR in the delayed system was reported by Yang and Liu [23-25]. Different from the classic theory, VR can be effectively controlled by adjusting the delay parameter instead of modulating the high-frequency signal. As an extension of VR, Yao and Zhan found that the weak low-frequency signal can propagate with high efficiency in one-way coupled bistable systems no matter the high-frequency signals added on all the oscillators or only on the first oscillator [26]. More interestingly, based on the VR mechanism, the signal propagation can be improved by an appropriate time delay feedback. The result was reported by Yang and Liu in globally delay-coupled oscillator chains [27].

In this paper, we investigate VR in a multistable system in both underdamped case

$$\ddot{x} + \delta\dot{x} + \frac{dV(x)}{dx} = f\cos(\omega t) + F\cos(\Omega t), \qquad (1)$$

and overdamped case

$$\dot{x} + \frac{dV(x)}{dx} = f\cos(\omega t) + F\cos(\Omega t) \qquad (2)$$

respectively, where $f \ll 1$ and $\omega \ll \Omega$. $V(x)$ is a simple periodic potential function, and $V(x) = -\cos x$.

The systems can model not only a particle moves in a multistable system but also a variety of other different physical mechanisms such as physical pendulums, Josephson junctions, phase-locking loops, *etc.* [28]. Here, we simply view the models as a weak low-frequency signal transmits in the one-dimensional multistable system.

The highlight of this paper is the finding of the sequential VR. At first, the theoretical investigation will be conducted following the method of direct separation of motions [29-32]. Then, the sequential VR will be discussed by both analytical and numerical results. The role of the damping coefficient is a focus in the underdamped system. Finally, conclusions will be given to conclude the paper.

## 2. Underdamped system

In this section, we investigate VR in system (1) through the method of direct separation of motions, which is widely used in obtaining the response amplitude of the output. First, we seek an approximate solution of equation (1) in the form

$$x(t) = X(t) + \Psi(t). \qquad (3)$$

$X(t)$ is a slowly varying term with frequency $\omega$, $\Psi(t)$ is a variable describing the fast motion which possesses of the period of $2\pi/\Omega$, *i.e.*,

$$\Psi(t) = \Lambda \cos(\Omega t + \phi), \qquad (4)$$

where $\Lambda = F/(\Omega\sqrt{\Omega^2 + \delta^2})$, $\sin\phi = -\delta/\sqrt{\Omega^2 + \delta^2}$, $\cos\phi = -\Omega/\sqrt{\Omega^2 + \delta^2}$. Substituting equation (4) into equation (1) and then averaging the equation over the time interval $[0, 2\pi/\Omega]$, we obtain the reduced equation on the slow variable $X(t)$, *i.e.*,

$$\ddot{X} + \delta \dot{X} + \frac{\mathrm{d}\overline{V}(X)}{\mathrm{d}X} = f\cos(\omega t). \qquad (5)$$

In equation (5), $\bar{V}(X) = -J_0(\Lambda)\cos X$ is the effective potential. Here, in the process of obtaining equation (5), the identities $\int_0^{2\pi/\Omega} \sin\Psi \, dt = 0$ and $\int_0^{2\pi/\Omega} \cos\Psi \, dt = \frac{2\pi}{\Omega} J_0(\Lambda)$ are used, where $J_0(\cdot)$ is the zero-order Bessel function of the first kind [33].

Let $X^*$ denotes the stable equilibrium of system (1), where $X^* = \pm 2n\pi$ and $n$ is an integer. Substituting $Y = X - X^*$ into equation (5), in the vicinity of $X^*$, it yields the linearization equation

$$\ddot{Y} + \delta \dot{Y} + J_0(\Lambda) Y = f \cos(\omega t). \tag{6}$$

Corresponding to system (6), the renormalized resonant frequency $\omega_r$ satisfies

$$\omega_r = \sqrt{|J_0(\Lambda)|}. \tag{7}$$

In the limit $t \to \infty$, the solution of equation (6) is $A_L \cos(\omega t - \varphi)$, where

$$A_L = \frac{f}{\sqrt{(\omega_r^2 - \omega^2)^2 + (\delta\omega)^2}}, \quad \varphi = \tan^{-1}\left(\frac{\omega_r^2 - \omega^2}{\delta\omega}\right). \tag{8}$$

As a result of above, the analytical expression for the response amplitude $Q$ is obtained as

$$Q = \frac{A_L}{f} = \frac{1}{\sqrt{(\omega_r^2 - \omega^2)^2 + \delta^2 \omega^2}}. \tag{9}$$

To verify the theoretical result, we numerically evaluate the response amplitude of the system (*i.e.*, the component from the Fourier spectrum) at the low-frequency $\omega$, which is given by

$$Q = \sqrt{Q_{\sin}^2 + Q_{\cos}^2}/f \tag{10}$$

with

$$Q_{\sin} = \frac{2}{kT}\int_0^{kT} x(t)\sin(\omega t)dt, \quad Q_{\cos} = \frac{2}{kT}\int_0^{kT} x(t)\cos(\omega t)dt, \tag{11}$$

where $T = 2\pi/\omega$ and $k$ is a positive integer. In the following simulations, without special explanation, the parameters are chosen as $f = 0.1$, $\omega = 0.1$, $\Omega = 3$, $k = 200$.

Let $S = \omega_r^2 - \omega^2$ and $W = (\omega_r^2 - \omega^2)^2 + (\delta\omega)^2$, from equation (9), it is evident that $Q$ achieves its maximum and the resonance appears if $S$ vanishes, which means the condition that $\omega_r=\omega$. According to the property of the zero-order Bessel function, there are sequential values of $F$, at which $\omega_r$ equals to $\omega$ ($W$ achieves its minimum), seeing figure 1. It leads to the appearances of a series of resonances, which is clearly shown in figure 2. In this paper, we named the phenomenon as a sequential VR, which is different from the traditional multiple VR that was reported by some researchers in the previous references [1, 10, 11, 21, 22]. Both the multiple VR and the sequential VR are characterized by more than one peak in the response amplitude $Q$, however the resonance in the sequential VR phenomenon seems much more regular, and the regularity is due to the inclusion of the zero-order Bessel function on the renormalized resonant frequency $\omega_r$. Also in figure 2, another fact is the agreement between the analytical results and the numerical simulations is influenced by the damping coefficient. For the case of small damping coefficient, as shown in figure 2 (a) and figure 2 (b), the resonant regions computed by the theoretical method are a little narrower than those obtained by the numerical simulation. A more interesting fact is that, according to the numerical results, the resonance occurs in some discrete regions that are a series of broad intervals on the $F$ axis, and this character also distinguishes from the traditional multiple VR. From equation (9), if the damping parameter is small, it is easy to understand the appearances of the discrete resonant regions. In the vicinity of $\omega_r=\omega$, the function $W$ approaches to a trivial value in some broad intervals of $F$. For example, in figure 1 (b), there are a series of discrete intervals in $F$-axis, at which the values of $W$ is very small, seeing that $W$ is below the horizontal line. These results directly lead to the fact that $Q$ reaches large values for a series of wide parameter intervals on $F$-axis. Accordingly, the numerical resonant regions in figure 2 (a) and figure 2 (b) appear. As to the analytical plots, the curves are bimodal in each resonant region. It is because $S=0$ has two positive roots in the considered interval. For example, corresponding to the first resonant region in figure 2 (a), the inset in figure 1 (a) shows that $F_1$ and $F_2$ are two different roots to the equation $S=0$. For a little bigger damping coefficient, the theoretical results are in good agreement with the results of numerical calculations, as shown in figure 2 (c) and figure 2 (d). In these two subplots, $Q$ varies gradually with the increase of $F$, the reason of which is that the value of $W$ varies quickly by adjusting the value of $F$, even though in the vicinity of $\omega_r=\omega$, *e.g.*, the curve for $\delta=1$ in figure 1 (b). According to above results, both the theoretical analysis and the numerical results are feasible, and both of them disclose the phenomenon of sequential VR, which is the highlight in this paper.

Figure 3 reveals the behavior of the response when $F$ is chosen in the resonant regions. In figure 3 (a), the response trajectories labeled by 1-4 corresponding to different $F$ in the resonant regions in figure 2 (a), it indicates that, when the resonance occurs, the particle may move into some specific wells (trajectory 1),or to far wells along one direction (trajectory 2-4). The different trajectories can be explained from the following viewpoint, *i.e.*, the mean drift velocity of the particle is directly dependent on $F$. Corresponding to trajectory 1-4, the mean drift velocity is zero, small and negative, big and positive, big and negative, respectively. Numerous references have discussed the related subject, especially in the ratchet device that under the biharmonic forces [30-32, 34]. In those references, the focus is the mobility of the particle, and it is characterized by the mean drift velocity. However, for the investigation on the VR phenomenon, the focus is the enhancement of the signal and the mobility of the particle is neglected. VR and mobility are two different themes in the vibrational mechanisms. Hence, here, we do not spend too much time to verify the dependence of the mean drift velocity with the value of $F$.  In figure 3, the subplots (b)-(e) give the trajectory in a short time interval, within which, the slow motion $X(t)$ is clearly depicted. According to figure 3, we know that the amplification of the signal is due to the response amplitude of the slow variable that contained in the output is much bigger than *f*, and the propagation of the signal is due to the last location of the motion is far from the initial potential well. The signal can propagate only when the resonance happens and the mean drift velocity is nontrivial.

In figure 4, we investigate the effect of the low-frequency $\omega$ on the sequential VR by using the numerical simulations. As shown in the figure, with the decrease of $\omega$, the response amplitude *Q* increases rapidly. In other words, improving signal via VR is especially significantly for the very-low-frequency signal. This fact also can be obtained from equation (9). It is because *Q* is a decreasing function of the variable $\omega$ when $\omega$ varies in the vicinity of $\omega_r$. In this figure, the other fact is that the resonant region turns wider with the increase of *F*, and which is easily to be understood from figure 1 (b). According to the property of the zero-order Bessel function, if we label $j_i$ as the *i*th (*i*=1, 2, ….) root of the equation $J_0(x)=0$, we then have $j_{n+1}-j_n=\pi$ if *n* is big enough. Hence, the spacing between the successive resonant regions becomes narrower with the increase of *F*.

## 3. Overdamped system

In this section, we pay our attention to system (2), and the method of direct separation of motions is still be used. Here, the expression of $\Psi(t)$ in equation (4) should be replaced by

$$\Psi(t) = \frac{F}{\Omega}\cos(\Omega t). \tag{12}$$

As the process in Section 2, it is easy to obtain the equation for the slow variable *X*, *i.e.*,

$$\dot{X} + \frac{\mathrm{d}\overline{V}(X)}{\mathrm{d}X} = f\cos(\omega t). \tag{13}$$

The effective potential $\bar{V}(X) = -J_0(F/\Omega)\cos X$. Substituting $Y=X-X^*$ into equation (13) and linearizing it in the vicinity of the stable state $X^*=\pm 2n\pi$, the linearization equation is obtained as

$$\dot{Y} + J_0(F/\Omega)Y = f\cos(\omega t) \tag{14}$$

Hence, the response amplitude $Q$ satisfies

$$Q = \frac{1}{\sqrt{\omega_r^2 + \omega^2}}, \tag{15}$$

where $\omega_r = |J_0(F/\Omega)|$.

The renormalized resonant frequency $\omega_r$ is a function of the zero-order Bessel function, its curve is plotted in figure 5 (a). In view of equation (15), we know the resonance occurs when $\omega_r$ achieves its minimum. From figure 5 (a), we see the minimum of $\omega_r$ recurrence with the increase of $F$. It results in the sequential VR in system (2), as is shown in figure 5 (b). Figure 6 gives the role of the low-frequency $\omega$ on the VR effect. The conclusion is the same as that in the underdamped system, *i.e.*, the VR is enhanced with the decrease of the low-frequency $\omega$. In these two figures, the analytical results match the numerical simulations closely.

## 4. Conclusions

In the present work, the phenomenon of sequential VR is reported in a multistable system in both underdamped and overdamped cases respectively. By applying the method of direct separation of motions, the analytical expressions of the response amplitude which is used to characterize the VR effect are obtained, and the theoretical results are in good agreement with the numerical ones. Because the zero-order Bessel function is included in the renormalized resonant frequency, it results in sequential appearances of VR via adjusting the amplitude of the excitation-including high-frequency signal. Further, in the underdamped system with a small damping coefficient, the VR occurs in a series of discrete broad parameter intervals, and the resonant regions are very apparently. The sequential VR distinguishes from the traditional multiple VR, and its appearance is much more regular with adjusting the high-frequency signal.

Finally, we introduce a general potential to extend the sequential VR in the multistable system, *i.e.*, $\overline{V}(x) = -\cos(nx)$. Following the schemes given in sections 2 and 3, for this general potential, we obtain the renormalized resonant frequency and the response amplitude

$$\omega_r = n\sqrt{|J_0(n\Lambda)|}, \quad Q = \frac{1}{\sqrt{(\omega_r^2 - \omega^2)^2 + \delta^2 \omega^2}} \quad (16)$$

and

$$\omega_r = n|J_0(nF/\Omega)|, \quad Q = \frac{1}{\sqrt{\omega_r^2 + \omega^2}} \quad (17)$$

for the underdamped case and overdamped case respectively. In the process of using the method of direct separation of motions to deduce the general results, the parameter $n$ should satisfy $n \ll \Omega/\omega$, and which assures the response includes a slow motion and a fast motion. In view of equations (16) and (17), the sequential VR phenomenon corresponding to this general potential can realize easily. Moreover, the effect of the parameter $n$, which determines the width of the potential, on the vibrational resonance can also be got directly. The result in this paper is a further development of VR, and it may be helpful in the field of signal processing electronics, especially for dealing with the very-low-frequency signal.

## Acknowledgments

The authors are very grateful to the anonymous reviewers for comments and suggestions. This research is supported by the National Science Foundation of China (Grant No. 11072107, 91016022) and the Specialized Research Fund for the Doctoral Program of Higher Education of China (Grant No. 20093218110003).

Figure captions

**Figure 1.** In the underdamped system, (a) The function of *S vs. F*. Inset: the enlarged view in the elliptical region. (b) The function of *W vs. F*.

**Figure 2.** The response amplitude *Q vs. F* presents sequential VR. (a) $\delta=0.05$, (b) $\delta=0.1$, (c) $\delta=0.5$, (d) $\delta=1.0$. Solid lines are theoretical results while dotted lines are numerically computed values.

**Figure 3.** (a) The trajectories in long time interval. (b)-(e) The trajectory in short time interval. The values of *F* are 21, 21.8, 22, 50.35, respectively, in subplots (b)-(e). The damping parameter $\delta=0.05$.

**Figure 4.** The effect of the low-frequency $\omega$ on the response amplitude *Q* in the underdamped system, $\delta=0.05$. From bottom to top, $\omega$ is chosen as 0.15, 0.1, 0.05 and 0.02 respectively.

**Figure 5.** In the overdamped system, (a) The resonant frequency $\omega_r$ *vs. F*. (b) The response amplitude *Q vs. F* presents sequential VR. Solid lines are theoretical results while the dots are numerically computed values.

**Figure 6.** The effect of low-frequency $\omega$ on the response amplitude *Q* in the overdamped system. From bottom to top, $\omega$ is chosen as 0.2, 0.15, 0.1 and 0.05 respectively. Solid lines are theoretical results while the dots are numerically computed values.

Figures

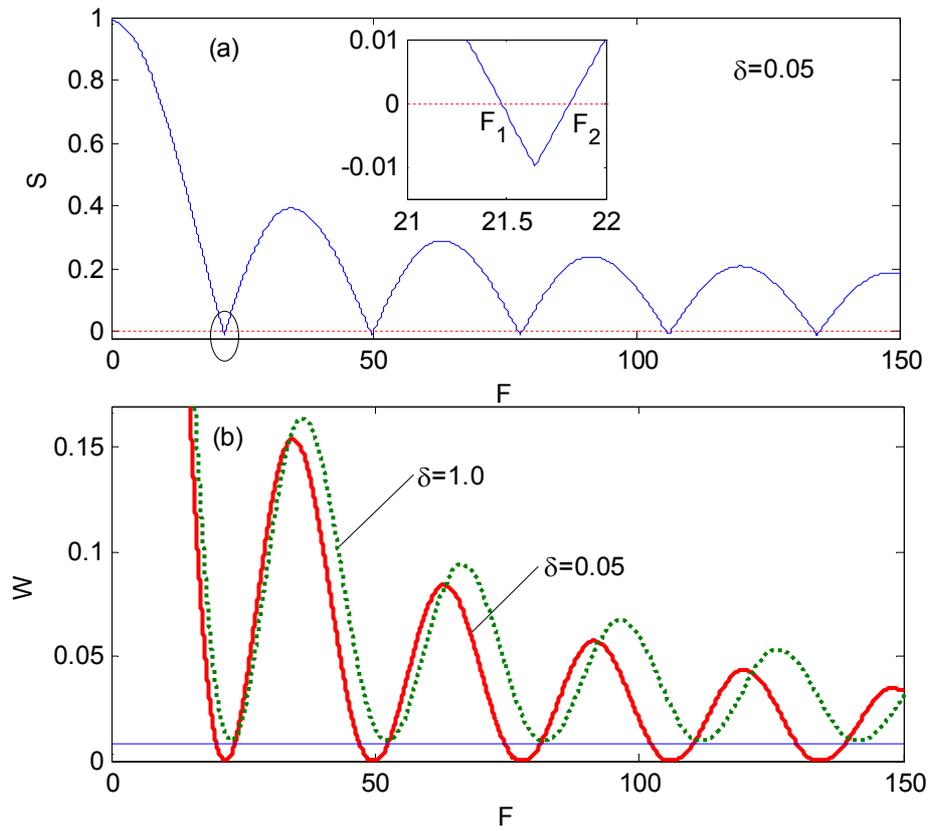

**Figure 1.** In the underdamped system, (a) The function of *S vs. F*. Inset: the enlarged view in the elliptical region. (b) The function of *W vs. F*.

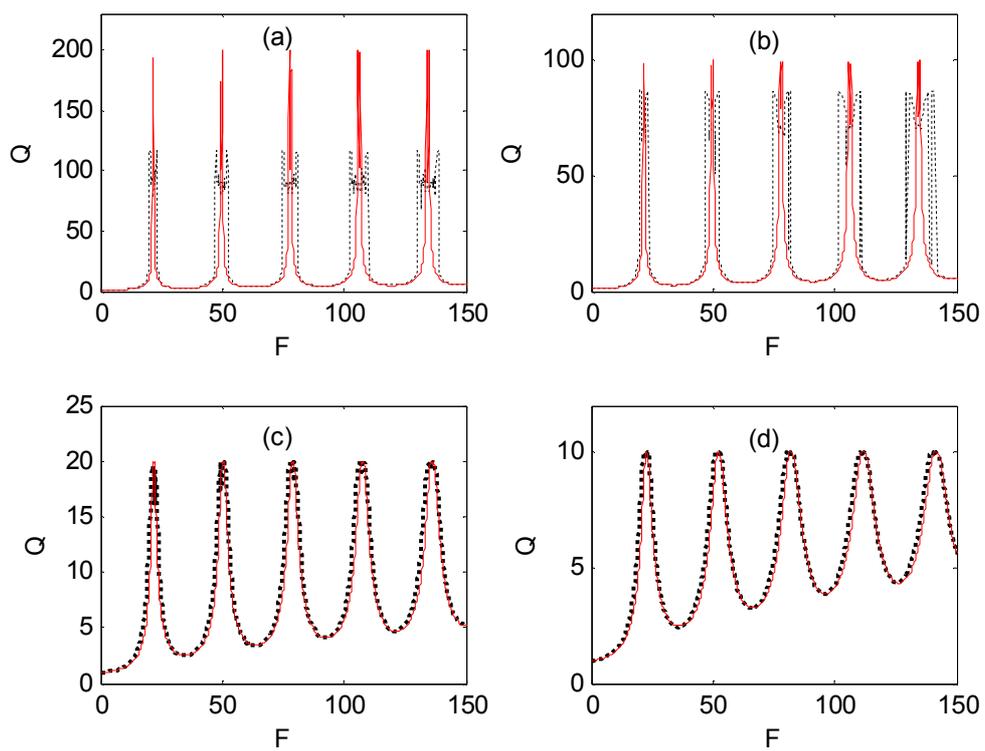

**Figure 2.** The response amplitude *Q vs. F* presents sequential VR. (a) $\delta$=0.05, (b) $\delta$=0.1, (c) $\delta$=0.5, (d) $\delta$=1.0. Solid lines are theoretical results while dotted lines are numerically computed values.

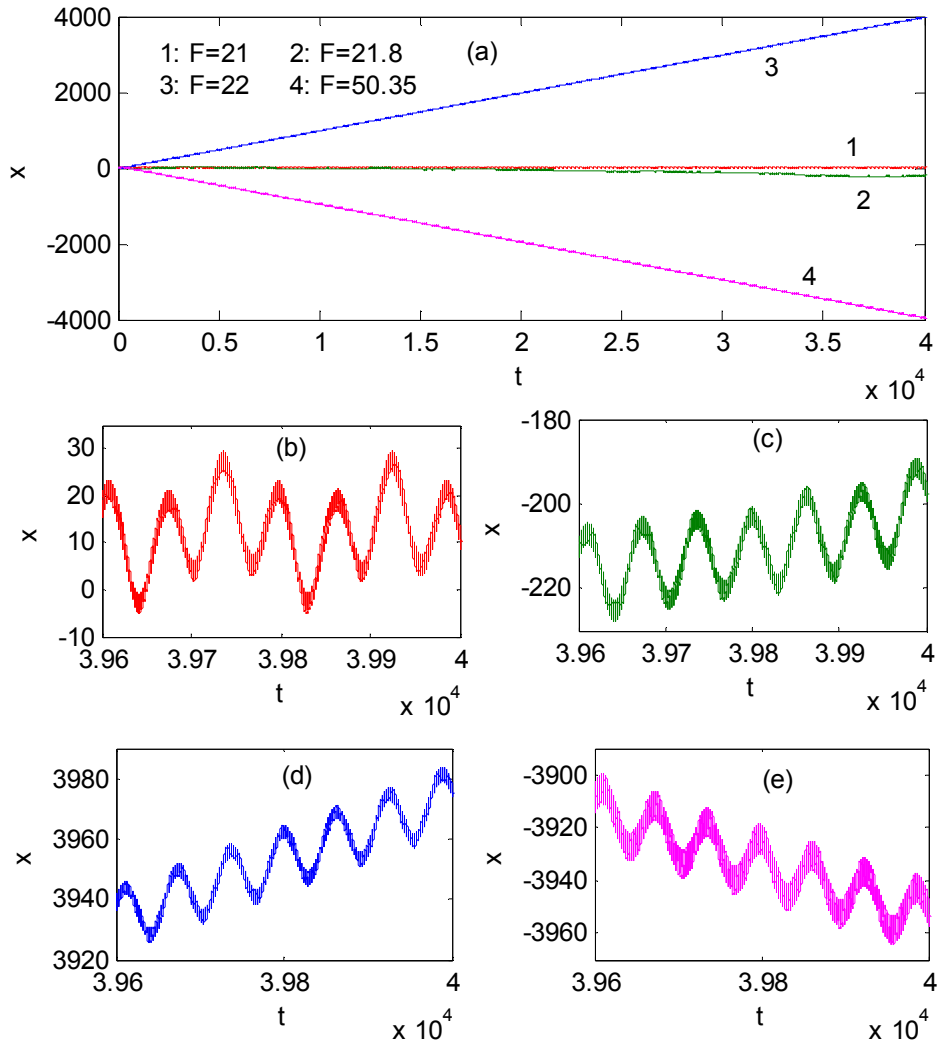

**Figure 3.** (a) The trajectories in long time interval. (b)-(e) The trajectory in short time interval. The values of *F* are 21, 21.8, 22, 50.35, respectively, in subplots (b)-(e). The damping parameter $\delta$=0.05.

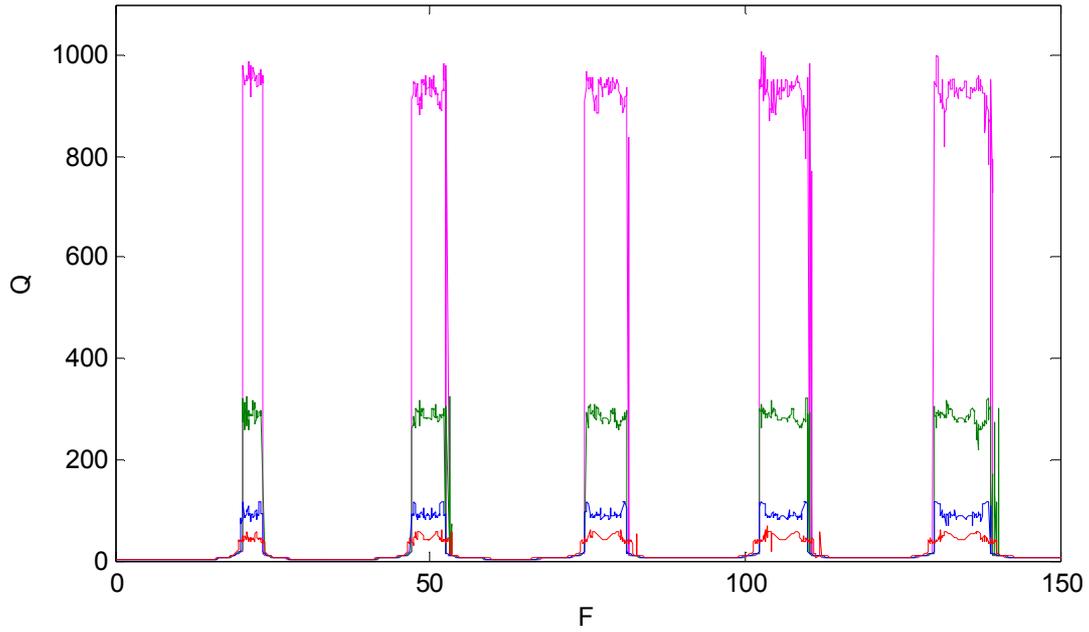

**Figure 4.** The effect of the low-frequency $\omega$ on the response amplitude $Q$ in the underdamped system, $\delta$=0.05. From bottom to top, $\omega$ is chosen as 0.15, 0.1, 0.05 and 0.02 respectively.

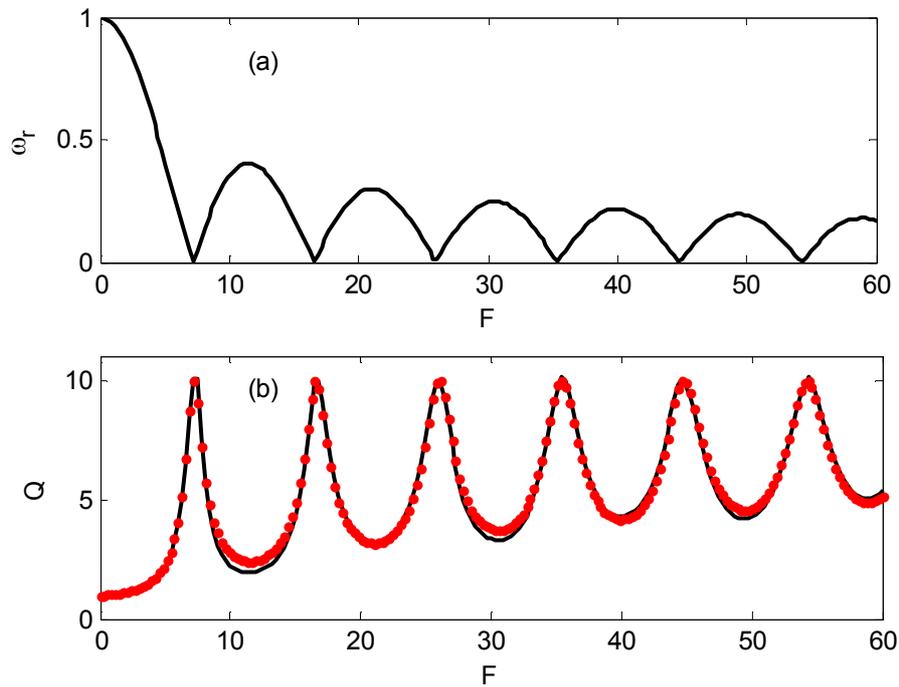

**Figure 5.** In the overdamped system, (a) The resonant frequency $\omega_r$ vs. $F$. (b) The response amplitude $Q$ vs. $F$ presents sequential VR. Solid lines are theoretical results while the dots are numerically computed values.

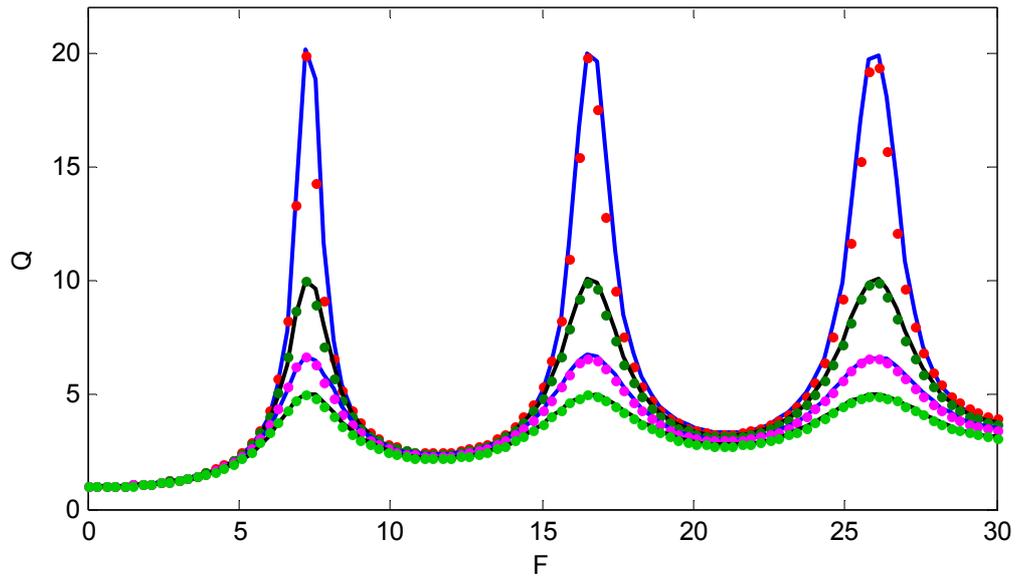

**Figure 6.** The effect of low-frequency $\omega$ on the response amplitude $Q$ in the overdamped system. From bottom to top, $\omega$ is chosen as 0.2, 0.15, 0.1 and 0.05 respectively. Solid lines are theoretical results while the dots are numerically computed values.